\documentclass[twocolumn,iop,numberedappendix,appendixfloats]{oja}

\usepackage[utf8]{inputenc}
\usepackage[british]{babel}

\usepackage{graphicx}	
\usepackage{amsmath}	
 \usepackage{amssymb}	
\usepackage{subcaption}  
\usepackage{caption}
\usepackage{xcolor}
\definecolor{maroon}{cmyk}{0, 0.87, 0.68, 0.32}
\definecolor{halfgray}{gray}{0.55}
\definecolor{ipython_frame}{RGB}{207, 207, 207}
\definecolor{ipython_bg}{RGB}{247, 247, 247}
\definecolor{ipython_red}{RGB}{186, 33, 33}
\definecolor{ipython_green}{RGB}{0, 128, 0}
\definecolor{ipython_cyan}{RGB}{64, 128, 128}
\definecolor{ipython_purple}{RGB}{170, 34, 255}
\usepackage{float}
\usepackage{fontawesome}
\usepackage[colorlinks,allcolors=blue]{hyperref}
\usepackage{url}

\newlength{\twocolwidth}

\usepackage{listings}

\lstdefinelanguage{iPython}{
    morekeywords={access,and,break,class,continue,def,del,elif,else,except,exec,finally,for,from,global,if,import,in,is,lambda,not,or,pass,print,raise,return,try,while},%
    morekeywords=[2]{abs,all,any,basestring,bin,bool,bytearray,callable,chr,classmethod,cmp,compile,complex,delattr,dict,dir,divmod,enumerate,eval,execfile,file,filter,float,format,frozenset,getattr,globals,hasattr,hash,help,hex,id,input,int,isinstance,issubclass,iter,len,list,locals,long,map,max,memoryview,min,next,object,oct,open,ord,pow,property,range,raw_input,reduce,reload,repr,reversed,round,set,setattr,slice,sorted,staticmethod,str,sum,super,tuple,type,unichr,unicode,vars,xrange,zip,apply,buffer,coerce,intern},%
    sensitive=true,%
    morecomment=[l]\#,%
    morestring=[b]',%
    morestring=[b]",%
    morestring=[s]{'''}{'''},
    morestring=[s]{"""}{"""},
    morestring=[s]{r'}{'},
    morestring=[s]{r"}{"},%
    morestring=[s]{r'''}{'''},%
    morestring=[s]{r"""}{"""},%
    morestring=[s]{u'}{'},
    morestring=[s]{u"}{"},%
    morestring=[s]{u'''}{'''},%
    morestring=[s]{u"""}{"""},%
    literate=
    {á}{{\'a}}1 {é}{{\'e}}1 {í}{{\'i}}1 {ó}{{\'o}}1 {ú}{{\'u}}1
    {Á}{{\'A}}1 {É}{{\'E}}1 {Í}{{\'I}}1 {Ó}{{\'O}}1 {Ú}{{\'U}}1
    {à}{{\`a}}1 {è}{{\`e}}1 {ì}{{\`i}}1 {ò}{{\`o}}1 {ù}{{\`u}}1
    {À}{{\`A}}1 {È}{{\'E}}1 {Ì}{{\`I}}1 {Ò}{{\`O}}1 {Ù}{{\`U}}1
    {ä}{{\"a}}1 {ë}{{\"e}}1 {ï}{{\"i}}1 {ö}{{\"o}}1 {ü}{{\"u}}1
    {Ä}{{\"A}}1 {Ë}{{\"E}}1 {Ï}{{\"I}}1 {Ö}{{\"O}}1 {Ü}{{\"U}}1
    {â}{{\^a}}1 {ê}{{\^e}}1 {î}{{\^i}}1 {ô}{{\^o}}1 {û}{{\^u}}1
    {Â}{{\^A}}1 {Ê}{{\^E}}1 {Î}{{\^I}}1 {Ô}{{\^O}}1 {Û}{{\^U}}1
    {œ}{{\oe}}1 {Œ}{{\OE}}1 {æ}{{\ae}}1 {Æ}{{\AE}}1 {ß}{{\ss}}1
    {ç}{{\c c}}1 {Ç}{{\c C}}1 {ø}{{\o}}1 {å}{{\r a}}1 {Å}{{\r A}}1
    {€}{{\EUR}}1 {£}{{\pounds}}1
    {^}{{{\color{ipython_purple}\^{}}}}1
    {=}{{{\color{ipython_purple}=}}}1
    {+}{{{\color{ipython_purple}+}}}1
    {-}{{{\color{ipython_purple}-}}}1
    {*}{{{\color{ipython_purple}$^\ast$}}}1
    {/}{{{\color{ipython_purple}/}}}1
    {+=}{{{+=}}}1
    {-=}{{{-=}}}1
    {*=}{{{$^\ast$=}}}1
    {/=}{{{/=}}}1,
    literate=
    *{-}{{{\color{ipython_purple}-}}}1
     {?}{{{\color{ipython_purple}?}}}1,
    identifierstyle=\color{black}\ttfamily,
    commentstyle=\color{ipython_cyan}\ttfamily,
    stringstyle=\color{ipython_red}\ttfamily,
    keepspaces=true,
    showspaces=false,
    showstringspaces=false,
    rulecolor=\color{ipython_frame},
    frameround={t}{t}{t}{t},
    numbers=none,
    numberstyle=\tiny\color{halfgray},
    backgroundcolor=\color{ipython_bg},
    basicstyle=\ttfamily\footnotesize,
    columns=fullflexible,
    keywordstyle=\color{ipython_green}\ttfamily,
}

\usepackage{array,multirow,graphicx}

\newcommand{\orcid}[1]{\href{https://orcid.org/#1}{\includegraphics[width=10pt]{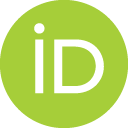}}}

\begin{document}

\journalinfo{The Open Journal of Astrophysics}
\submitted{submitted XXX; accepted YYY}

\title{\texttt{CosmoPower-JAX}: high-dimensional Bayesian inference\\with differentiable cosmological emulators\vspace{-10ex}}

\shorttitle{\texttt{CosmoPower-JAX}}
\shortauthors{D. Piras \& A. Spurio Mancini}

\author{D. Piras$^{\star1}$\orcid{0000-0002-9836-2661}} 
\author{A. Spurio Mancini$^{\dagger2}$\orcid{0000-0001-5698-0990}}

\affiliation{$^1$ Département de Physique Théorique, Université de Genève, 24 quai Ernest Ansermet, 1211 Genève 4, Switzerland}
\affiliation{$^2$ Mullard Space Science Laboratory, University College London, Holmbury St. Mary, Dorking, Surrey, RH5 6NT, UK}

\thanks{$^\star$ E-mail: \nolinkurl{davide.piras@unige.ch}}
\thanks{$^\dagger$ E-mail: \nolinkurl{a.spuriomancini@ucl.ac.uk}}

\date{\today}

\begin{abstract}
We present \texttt{CosmoPower-JAX}, a \texttt{JAX}-based implementation of the \texttt{CosmoPower} framework, which accelerates cosmological inference by building neural emulators of cosmological power spectra. We show how, using the automatic differentiation, batch evaluation and just-in-time compilation features of \texttt{JAX}, and running the inference pipeline on graphics processing units (GPUs), parameter estimation can be accelerated by orders of magnitude with advanced gradient-based sampling techniques. These can be used to efficiently explore high-dimensional parameter spaces, such as those needed for the analysis of next-generation cosmological surveys. We showcase the accuracy and computational efficiency of \texttt{CosmoPower-JAX} on two simulated Stage IV configurations. We first consider a single survey performing a cosmic shear analysis totalling 37 model parameters. We validate the contours derived with \texttt{CosmoPower-JAX} and a Hamiltonian Monte Carlo sampler against those derived with a nested sampler and without emulators, obtaining a speed-up factor of $\mathcal{O}(10^3)$. We then consider a combination of three Stage IV surveys, each performing a joint cosmic shear and galaxy clustering (3x2pt) analysis, for a total of 157 model parameters. Even with such a high-dimensional parameter space, \texttt{CosmoPower-JAX} provides converged posterior contours in 3 days, as opposed to the estimated 6 years required by standard methods. \texttt{CosmoPower-JAX} is fully written in \texttt{Python}, and we make it publicly available to help the cosmological community meet the accuracy requirements set by next-generation surveys. \href{https://github.com/dpiras/cosmopower-jax}{\faicon{github}}
\looseness=-1
\end{abstract}


\maketitle



\section{Introduction}
Bayesian inference of cosmological parameters from next-generation large-scale structure (LSS) and cosmic microwave background (CMB) surveys such as \textit{Euclid} \citep{Laureijs11}\footnote{\href{https://www.euclid-ec.org/}{https://www.euclid-ec.org/}}, the Vera Rubin Observatory \citep{Ivezic19}\footnote{\href{https://www.lsst.org/}{https://www.lsst.org/}}, the Nancy Grace Roman Space Telescope \citep{Spergel15}\footnote{\href{https://roman.gsfc.nasa.gov/}{https://roman.gsfc.nasa.gov/}}, the Simons Observatory \citep{Ade19}\footnote{\href{https://simonsobservatory.org/}{https://simonsobservatory.org/}}, CMB-S4 \citep{Abazajian16}\footnote{\href{https://cmb-s4.org/}{https://cmb-s4.org/}}, and CMB-HD \citep{Sehgal19}\footnote{\href{https://cmb-hd.org/}{https://cmb-hd.org/}}, will require the exploration of high-dimensional parameter spaces -- $\mathcal{O}(100)$ parameters and higher -- necessary to accurately model the physical signals and their several systematic contaminants. Sampling the posterior distribution in these high-dimensional spaces represents a significant computational challenge for Markov Chain Monte Carlo (MCMC) algorithms \citep{Roberts97, Katafygiotis08, Liu09}, which are traditionally used in cosmological analyses \citep{Lewis02, Audren12, Brinckmann18, Torrado21}. Gradient-based inference methods, such as Hamiltonian Monte Carlo (HMC, \citealp{Duane87, Neal96}) and variational inference (VI, \citealp{, Hoffman13, Blei17}), manage to concentrate the sampling in regions of high posterior mass, even in large parameter spaces, provided one has efficient access to accurate derivatives of the likelihood function with respect to the model parameters \citep{Brooks11, Neal11, Zhang17, Betancourt17}. Combining differentiable and computationally inexpensive likelihood functions with gradient-based inference techniques is thus crucial to efficiently obtain unbiased posterior distributions in high-dimensional parameter spaces.

Recently, there has been growing interest in the use of machine learning emulators to accelerate cosmological parameter estimation \citep[e.g.][]{Mootoovaloo20, Arico21, Zennaro21, Nygaard22, Gunther22, Bonici22, Eggemeier22}. 
\citet{CP} (SM22 hereafter), in particular, developed \texttt{CosmoPower}, a suite of neural network emulators of cosmological power spectra that replaces the computation of these quantities traditionally performed with Einstein-Boltzmann solvers such as the Code for Anisotropies in the Microwave Background (CAMB, \citealt{Lewis11}) or the Cosmic Linear Anisotropy Solving System (CLASS, \citealt{Blas11}). In SM22 the authors show how Bayesian inference of cosmological parameters can be accelerated by several orders of magnitude using \texttt{CosmoPower}; the speed-up becomes particularly relevant when the emulators are employed within an inference pipeline that can be run on graphics processing units (GPUs).

An additional advantage in using machine learning emulators is that they efficiently provide accurate derivatives with respect to their input parameters. This is made possible by the \textit{automatic differentiation} features implemented in the libraries routinely used to build these emulators, such as \texttt{TensorFlow} \citep{Abadi15}, \texttt{JAX} \citep{Jax18} or \texttt{PyTorch} \citep{Paszke19}. Automatic differentiation (\textit{autodiff}, \citealp{Bartholomew00, Neidinger10, Baydin18, Paszke19b, Margossian19}) can be used in gradient-based sampling algorithms to compute the derivatives of the likelihood function, provided the emulators are inserted within likelihood functions that can also be automatically differentiated. In practice, this means that the entire software implementation of the likelihoods should be written using primitives that belong to the aforementioned libraries. This is, for example, the main idea behind \texttt{jax-cosmo} \citep{Campagne23}, a \texttt{JAX}-based library recently developed to compute cosmological observables, validated against the widely used Core Cosmology Library (CCL, \citealp{Chisari19}). For other examples of gradient-based inference in cosmology, see e.g. \citet{Hajian07, Taylor08, Jasche10, Jasche10b, Lavaux15, Nguyen21, Valade22, Kostic22, Loureiro23, Porqueres23, Ruiz23}.

In this paper, we first develop an implementation of \texttt{CosmoPower} built using purely \texttt{JAX} primitives. We train the neural network emulators on the same data used in SM22, achieving an accuracy similar to that obtained by the \texttt{TensorFlow}-based implementation of SM22. We additionally compare the derivatives of the power spectra with respect to the input cosmological parameters computed using \textit{autodiff} against those obtained with numerical differences and CAMB, showing a good agreement overall. We then couple \texttt{CosmoPower-JAX} with \texttt{jax-cosmo} to build likelihood functions completely written using the \texttt{JAX} library; this allows us to run parameter estimation pipelines using gradient-based algorithms on GPUs over very large parameter spaces. In particular, we consider two examples of future Stage IV survey configurations. In the first one, we study a single cosmic shear survey with typical Stage IV specifications; we compare the posterior contours obtained with our \texttt{JAX}-based pipeline against those obtained with a standard pipeline based on \texttt{CCL}, CAMB and the nested sampler \texttt{PolyChord} \citep{Handley15, Handley15b}. In the second case, we consider a joint analysis of three Stage IV surveys, each carrying out a joint shear - galaxy clustering (3x2pt) analysis. We show how \texttt{CosmoPower-JAX} enables efficient exploration of the 157-dimensional parameter space required for the modelling of the observables at a fraction of the time that would be required with traditional methods. 
In both experiments, inference is performed using the No U-Turn Sampler (NUTS) HMC variation \citep{Hoffman14}, as implemented in the \texttt{NumPyro} library \citep{Bingham19, Phan19}.

The structure of this paper is as follows. In Sec.~\ref{sec:methods} we first describe our new \texttt{JAX} emulators and how they are trained on the same data used in SM22. We then review gradient-based methods for posterior inference and describe our \texttt{JAX}-based cosmological likelihoods. In Sec.~\ref{sec:results} we validate the accuracy of our emulators and report our inference results in the two simulated Stage IV configurations described above. We conclude in Sec.~\ref{sec:conclusions}. We make \texttt{CosmoPower-JAX}, which is fully written in \texttt{Python}, publicly available.\footnote{\href{https://github.com/dpiras/cosmopower-jax}{https://github.com/dpiras/cosmopower-jax}}

\section{Methods}\label{sec:methods}
\subsection{\normalfont\texttt{CosmoPower-JAX}}
\texttt{CosmoPower} is a suite of neural network emulators of cosmological power spectra, namely matter and cosmic microwave background power spectra (albeit the emulation framework is highly flexible and can be used to emulate any other cosmological quantity; \citealp[see e.g.][]{Burger23, Gong23}). Power spectra represent the key component of 2-point statistics analysis of cosmological fields, and are typically computed by Einstein-Boltzmann solvers such as CAMB or CLASS. These computations usually represent the main bottleneck in parameter estimation pipelines, particularly when high accuracy is required or when the cosmological scenario considered is not standard (see e.g. \citealp{Spurio22, Balkenhol22, Bolliet23}).

\texttt{CosmoPower-JAX} is a \texttt{JAX}-based implementation of \texttt{CosmoPower}; as such, it implements the same two emulation methods present in the original \texttt{CosmoPower} software. These are either a direct neural network mapping between cosmological parameters and logarithmic spectra, or a neural network mapping between cosmological parameters and the coefficients of a principal component analysis (PCA) of the spectra; we refer to SM22 for a detailed discussion. In our release version of \texttt{CosmoPower-JAX} these differences are dealt with internally; therefore, the user can simply obtain fast predictions of cosmological power spectra in three lines of code. 

The neural networks we employ are made of four layers of 512 nodes each, with the same activation function described in SM22. In the case of the CMB temperature-polarisation and lensing power spectra, we preprocess the spectra using PCA, keeping 512 and 64 components, respectively. We optimise the mean-squared-error loss function using \texttt{Adam} \citep{Kingma15}, and keep a constant batch size of 512. The starting learning rate is 10$^{-2}$, which we decrease by a factor of 10 (until 10$^{-6}$) every time the validation loss does not improve for more than 40 consecutive epochs.

We choose to re-implement and re-train the \texttt{CosmoPower} emulators in \texttt{JAX} to demonstrate that it is possible to build neural networks with this library; nonetheless, in our public implementation of \texttt{CosmoPower-JAX} it is also possible to upload weights and biases from previously-trained models and obtain \texttt{JAX} predictions using only the forward pass of the network. Building the emulators using the \texttt{JAX} library unlocks its automatic-differentiation features, as well as the efficient batch evaluation and just-in-time compilation. Importantly, \texttt{JAX} is built around the popular \texttt{NumPy} library \citep{Harris20}, which facilitates its use and portability. We refer the reader to \citet{Jax18} and \citet{Campagne23} for a complete description of the features of \texttt{JAX}, especially in the context of cosmological analyses.


\subsection{Data}
We consider the same datasets as in SM22 to train and validate our emulators. Using CAMB, we produce a total of $\sim 2\times10^5$ matter power spectra at 420 wavenumbers in the interval $k \in \left[10^{-5} - 10 \right]$ Mpc$^{-1}$ and with varying redshift $z \in \left[0, 5\right]$. We consider both the linear matter power spectrum $P_{\delta \delta}^{\rm{L}}$ and the non-linear correction $P_{\delta \delta}^{\rm{NL-CORR}}$, such that the non-linear matter power spectrum $P_{\delta \delta}$ can be written as:
\begin{align}
    P_{\delta \delta} (k, z) = P_{\delta \delta}^{\textrm{L}} (k, z) \, P_{\delta \delta}^{\textrm{NL-CORR}} (k, z) \ ,
    \label{eq:pdd}
\end{align}
where $P_{\delta \delta}^{\textrm{NL-CORR}}$ is computed with \texttt{HMcode} \citep{Mead15, Mead16}. This non-linear correction introduces two extra baryonic parameters: the minimum halo concentration $c_{\textrm{min}}$, and the halo bloating $\eta_0$, which we vary when performing Bayesian inference. 

Using CAMB we also produce $\sim 5\times10^5$ CMB power spectra, including the temperature $C^{\textrm{TT}}_{\ell}$, polarisation $C^{\textrm{EE}}_{\ell}$, temperature-polarisation $C^{\textrm{TE}}_{\ell}$ and lensing potential $C^{\phi\phi}_{\ell}$power spectra in the interval $\ell \in \left[ 2, 2508 \right] $. All emulators are trained on the same parameter range indicated in table 1 of SM22. In both cases, we use $\sim 90$\% of the data to train the neural networks (of which we leave 10\% for validation), and evaluate the trained models on the remaining $\mathcal{O}(10^4)$ test spectra.

\subsection{Gradient-based Bayesian inference}
The goal of Bayesian inference is to sample the posterior distribution $p(\boldsymbol{\theta} | \boldsymbol{d})$ of some parameters $\boldsymbol{\theta}$ given observed data $\boldsymbol{d}$. Bayes' theorem relates the posterior distribution to the likelihood function $p(\boldsymbol{d} | \boldsymbol{\theta})$:
\begin{align}
    p(\boldsymbol{\theta} | \boldsymbol{d}) = \frac{p(\boldsymbol{d} | \boldsymbol{\theta}) p(\boldsymbol{\theta})}{p(\boldsymbol{d})},
\end{align}
where $p(\boldsymbol{\theta})$ expresses prior knowledge on the parameters $\boldsymbol{\theta}$, and the evidence $p(\boldsymbol{d})$ is a normalisation factor commonly ignored in parameter estimation tasks.

Drawing from $p(\boldsymbol{\theta} | \boldsymbol{d})$ requires the use of stochastic samplers to generate a list of samples drawn from the posterior distribution. MCMC and nested sampling algorithms are the two main classes of samplers typically used in cosmological applications. Widely used examples of MCMC methods in cosmology include the Metropolis-Hastings (with its variations) and affine ensemble algorithms \citep{Lewis02, Lewis13, Karamanis22b, Goodman10, ForemanMackey13}, whereas the most widely adopted nested sampling algorithms are multimodal nested sampling and slice-based nested sampling \citep{Feroz08, Feroz09, Feroz19, Handley15, Handley15b}.

Metropolis-Hastings and nested sampling methods can struggle to explore large parameter spaces, since, as the dimensionality increases, sampling from the typical set of the posterior distribution becomes exponentially harder \citep{Betancourt17}. In these challenging inference scenarios, it is useful to resort to gradient-based algorithms, which manage to concentrate the sampling in regions of high posterior mass despite the large number of model parameters. A popular gradient-based method is Hamiltonian Monte Carlo (HMC, \citealp{Duane87, Neal96}), which formulates the problem of sampling the posterior distribution as the dynamical evolution of a particle with position $\boldsymbol{\theta}$ and momentum $\boldsymbol{p}$. The Hamiltonian of the system is written as:
\begin{equation}
 \mathcal{H} (\boldsymbol{\theta}, \boldsymbol{p}) = \frac{1}{2} \boldsymbol{p}^T \mathcal{M}^{-1} \boldsymbol{p} + \textbf{U}(\boldsymbol{\theta}) \ ,
\end{equation}
where $\mathcal{M}$ is a ``mass matrix'', which is assumed to be the covariance matrix of a zero-mean multivariate Gaussian distribution used to sample $\boldsymbol{p}$, and the potential energy $\textbf{U}(\boldsymbol{\theta})$ is defined as:
\begin{equation}
    \textbf{U}(\boldsymbol{\theta}) \equiv -\ln p(\boldsymbol{\theta} | \boldsymbol{d}) \ .
\end{equation}
Starting from an initial point in the phase space $(\boldsymbol{\theta}, \boldsymbol{p})$, it is possible to collect samples of the posterior distribution by numerically solving the Hamilton's equations using a leapfrog algorithm. A new proposed state $(\boldsymbol{\theta}^*, \boldsymbol{p}^*)$ is then either accepted or rejected based on the new energy value $\mathcal{H} (\boldsymbol{\theta}^*, \boldsymbol{p}^*)$. HMC can thus be seen as a modification to the original Metropolis-Hastings algorithm, which leverages the analogy with a Hamiltonian system to find a better proposal distribution and results in a higher acceptance rate. 

A practical difficulty of HMC is the need for tuning of the hyperparameters governing the leapfrog numerical integration of the Hamiltonian dynamics. These are the number and the size of steps to be taken by the integrator before the sampler changes direction to a new random one. These numbers can be particularly hard to tune and lead to inefficient sampling of the posterior. The No U-Turn Sampler (NUTS, \citealp{Hoffman14}) tackles this issue by forcing the sampler to avoid U-turns in parameter space. In particular, rather than fixing the number of integration steps, NUTS adjusts the integration length by running the leapfrog algorithm until the trajectory starts to return to previously-visited regions of the parameter space; while the number of model evaluations increases, this leads to a higher acceptance rate, and therefore to faster and more reliable convergence.

In our experiments, we use the NUTS implementation provided by the \texttt{NumPyro} library \citep{Bingham19, Phan19}. We set the integration step size to $10^{-3}$, and find no difference in the final results by increasing or decreasing this value by one order of magnitude. Additionally, we set the maximum number of model evaluations before sampling a new momentum to $2^7$, and specify a block mass matrix according to the expected correlations among parameters. Finally, to further improve the geometry of the problem we perform a reparameterisation of the inference variables: all of the parameters with a Gaussian prior distribution are decentered \citep{Gorinova20}, while all of those with a uniform prior are rescaled to $\mathcal{U} \left[-5, 5, \right]$. All original prior distributions are reported in Table~\ref{tab:priors}.

We run the NUTS algorithm on three NVIDIA A100 GPUs with 80 GB of memory.
Using multiple GPUs allows us to showcase the \texttt{pmap} feature of \texttt{JAX} to distribute the NUTS chains over multiple GPU devices. We report below an example snippet to run inference over multiple GPUs using \texttt{NumPyro}, where the parallelisation happens simultaneously over a single device and across multiple devices.

\newpage

\begin{lstlisting}[language=iPython]
def do_mcmc(rng_key, n_vectorized=8):
  # All posterior details are defined in this kernel
  nuts_kernel = NUTS(log_posterior)
  mcmc = MCMC(
    nuts_kernel,
    num_chains=n_vectorized,
    chain_method="vectorized"
    )
  mcmc.run(rng_key)
  return {**mcmc.get_samples()}

# Number of devices to pmap over
n_parallel = jax.local_device_count()
rng_ks = jax.random.split(PRNGKey(rng_seed), n_parallel)
samples = pmap(do_mcmc)(rng_ks)

\end{lstlisting}

In this example, 8 chains are run on each GPU, and the total number of GPUs available is determined at run time.

\subsection{Likelihoods}
\label{sec:likelihoods}

\begin{table}
  \centering 
  \renewcommand{\arraystretch}{2}
  \setlength{\tabcolsep}{3.5pt}
  \begin{tabular}{c c c c}
    &\textbf{Parameter}               &  \textbf{Prior range}  & \textbf{Fiducial value} \\
    \hline
    \hline
     \parbox[c]{-5mm}{\multirow{5}{*}{\rotatebox[origin=c]{90}{\textbf{Cosmology}}}} 
     &$\omega_{\mathrm{b}} = \Omega_{\mathrm{b}} h^2$            & $\mathcal{U}$[0.01875, 0.02625]     & 0.02242\\
    &$\omega_{\mathrm{cdm}}= \Omega_{\mathrm{cdm}} h^2$          & $\mathcal{U}$[0.05, 0.255]          & 0.11933\\
    &$h$                              & $\mathcal{U}$[0.64, 0.82]           & 0.6766\\
    &$n_{\textrm{s}}$                            & $\mathcal{U}$[0.84, 1.1]            & 0.9665\\
    &$\mathrm{ln}10^{10}A_{\textrm{s}}$          & $\mathcal{U}$[1.61, 3.91]           & 3.047\\
    \hline
    \hline
    \\[-4ex] \parbox[c]{-5mm}{\multirow{2}{*}{\rotatebox[origin=c]{90}{\textbf{Baryons} }}} 
     &$c_{\mathrm{min}}$               & $\mathcal{U}$[2, 4]                 & 2.6\\
    &$\eta_0$                         & $\mathcal{U}$[0.5, 1]               & 0.7 \\[1.8ex]
    \hline
    \hline
    \parbox[c]{-5mm}{\multirow{5}{*}{\rotatebox[origin=c]{90}{\textbf{Nuisance S1}}}}& 
    $A^{\mathrm{S1}}_{\mathrm{IA}, i}$ \hfill $i=1, \dots, 10$                & $\mathcal{U}$[$-6$, 6]                & $1 - 0.1i$\\
    &$D^{\mathrm{S1}}_{z_i, \textrm{source}}  \hfill i = 1, \dots, 10$  & $\mathcal{N}(0, 10^{-4})$  & 0\\
    &$m^{\mathrm{S1}}_i$  \hfill $i=1, \dots, 10$                &$\mathcal{N}(0.01, 0.02)$               & 0.01\\
    &$D^{\mathrm{S1}}_{z_i, \textrm{lens}}  \hfill i = 1, \dots, 10$  & $\mathcal{N}(0, 10^{-4})$  & 0\\
    &$b^{\mathrm{S1}}_{i}$  \hfill  $i=1, \dots, 10$                & $\mathcal{U}$[0.1, 5]                & 1 \\
    \hline
    \hline
    \parbox[c]{-5mm}{\multirow{5}{*}{\rotatebox[origin=c]{90}{\textbf{Nuisance S2}}}}&$A^{\mathrm{S2}}_{\mathrm{IA}, i}$ \hfill$i=1, \dots, 10$                & $\mathcal{U}$[0, 10]                & $6.45 - 0.5i$\\
    &$D^{\mathrm{S2}}_{z_i, \textrm{source}} \hfill i = 1, \dots, 10$  & $\mathcal{N}(0, 10^{-3})$  & 0\\
    &$m^{\mathrm{S2}}_i$ \hfill $i=1, \dots, 10$                &$\mathcal{N}(0, 0.002)$               & 0\\
    &$D^{\mathrm{S2}}_{z_i, \textrm{lens}} \hfill i = 1, \dots, 10$  & $\mathcal{N}(0, 10^{-3})$  & 0\\
    &$b^{\mathrm{S2}}_{i}$ \hfill $i=1, \dots, 10$                & $\mathcal{U}$[0.8, 3]                & 1.2+0.1$i$ \\
    \hline
    \hline
    \parbox[c]{-5mm}{\multirow{5}{*}{\rotatebox[origin=c]{90}{\textbf{Nuisance S3}}}}    &$A^{\mathrm{S3}}_{\mathrm{IA}, i}$ \hfill $i=1, \dots, 10$                & $\mathcal{U}$[0, 6]                & $1.1 - 0.1i$\\
    &$D^{\mathrm{S3}}_{z_i, \textrm{source}} \hfill i = 1, \dots, 10$  & $\mathcal{N}(0, 10^{-3})$  & 0\\
    &$m^{\mathrm{S3}}_i$ \hfill$i=1, \dots, 10$                &$\mathcal{N}(0, 0.003)$               & 0\\
    &$D^{\mathrm{S3}}_{z_i, \textrm{lens}} \hfill i = 1, \dots, 10$  & $\mathcal{N}(0, 10^{-3})$  & 0\\
    &$b^{\mathrm{S3}}_{i}$ \hfill$i=1, \dots, 10$                & $\mathcal{U}$[0.8, 3]                & 1.25+0.05$i$ \\
    \hline
    \hline
    \end{tabular}
   \caption{Prior distributions and fiducial values of the cosmological and nuisance parameters for the simulated analyses performed in this work. Uniform (Gaussian) prior distributions are indicated with $\mathcal{U}$ ($\mathcal{N}$). For the cosmological parameters and the baryonic feedback parameters ($c_{\mathrm{min}}, \eta_0$) the prior range corresponds to the range of validity of our emulators. The superscripts S1, S2 and S3 refer to three simulated Stage IV surveys, whose details we discuss in Sect.~\ref{sec:likelihoods}.}
  \label{tab:priors}
\end{table}

To test the computational efficiency and accuracy of our \texttt{CosmoPower-JAX} emulators, we run examples of cosmological inference on large parameter spaces, such as those that will characterise the analysis of future surveys. We consider a power spectrum analysis of both cosmic shear alone, as well as in cross-correlation with galaxy clustering (3x2pt), which in recent years has become the standard method to combine information on the cosmic shear and galaxy clustering field \citep[e.g.][]{Joachimi10, Sanchez21, Heymans21, Abbott22}. Redshift information is included in the analysis through tomography, assigning galaxies to one of $N_{\rm bins}$ bins based on their photometric redshift. The angular power spectra of the probes are jointly modelled computing all of the unique correlations between tomographic bins $i, j = 1, \dots, N_{\rm bins}$. In the following we briefly review the modelling of the power spectra for cosmic shear, galaxy clustering and their cross-correlation (``galaxy-galaxy lensing''), as well as describe the Stage IV surveys we take into consideration. We follow the notation of SM22. 

Angular power spectra for the three probes can be expressed as integrals of the matter power spectrum $P_{\delta \delta}(k, z)$ (a function of wavevector $k$ and redshift $z$), weighted by pairs of window functions $W$ for the shear ($\gamma$), clustering ($\mathrm{n}$) and intrinsic alignment ($\mathrm{I}$) fields:
\begin{align}\label{eq:general_cell}
    C_{i j}^{A B}(\ell) = \int_0^{\chi_{\mathrm{H}}} \mathrm{d}\chi \, \frac{W_i^A \, W_j^B}{\chi^2} \, P_{\delta \delta} \left(k = \frac{\ell + 1/2}{\chi}, z \right),
\end{align}
where $\{ A ,B \} = \{ \gamma, \mathrm{n}, \mathrm{I} \}$, $\chi$ is the comoving distance (itself a function of redshift), and the upper limit of integration is the Hubble radius $\chi_{\rm H} = c/H_0$, with $c$ the speed of light and $H_0$ the Hubble constant. In Eq.~(\ref{eq:general_cell}) we assumed the extended Limber approximation \citep{LoVerde08} which connects Fourier scales $k$ with angular multipoles $\ell$.

The angular power spectrum $C_{ij}^{\epsilon \epsilon}(\ell)$ of the cosmic shear signal for tomographic bins $i$ and $j$ is a combination of a pure shear contribution and an intrinsic alignment contamination:
\begin{align}\label{eq:cell_shear}
C_{ij}^{\epsilon \epsilon}(\ell) = C_{ij}^{\gamma \gamma}(\ell) + C_{ij}^{\gamma \mathrm{I}}(\ell) + C_{ij}^{\mathrm{I} \gamma}(\ell) + C_{ij}^{\mathrm{I} \mathrm{I}}(\ell) \ .
\end{align}
For a tomographic redshift bin distribution $n_{i, , \textrm{source}}(z), i = 1, \dots, N_{\rm bins}$, the weighting function $W_i^\gamma (\chi)$ for the cosmic shear field $\gamma$ is given by:
\begin{align}\label{eq:window_shear}
    W_i^\gamma (\chi) = \frac{3 \, H_0^2 \, \Omega_{\rm m}}{2 \, c^2} \frac{\chi}{a} \int_{\chi}^{\chi_{\rm H}} \mathrm{d} \chi' \, n_{i, \textrm{source}}(\chi') \, \frac{\chi'-\chi}{\chi'} \ ,
\end{align}
where $\Omega_{\rm m}$ is the matter density parameter and $a$ is the scale factor. To model the window function for the intrinsic alignment field $\mathrm{I}$, we start from the commonly-used non-linear alignment model \citep{Hirata04, Joachimi11}: 
\begin{align}\label{eq:window_alignment_original}
    W_i^{\rm I}(\chi) = - A_{\rm IA} \left( \frac{1+z}{1+z_{\rm p}} \right)^{\eta_{\rm IA}} \frac{C_1 \, \rho_{\rm cr} \, \Omega_{\rm m}}{D(\chi)} \, n_{i, \textrm{source}}(\chi) \ ,
\end{align}
with two free parameters $A_{\rm IA}$ and $\eta_{\rm IA}$; the linear growth factor $D(\chi)$, the critical matter density $\rho_{\rm cr}$, a constant $C_1$ and a pivot redshift $z_{\rm p}$ also enter Eq.~(\ref{eq:window_alignment_original}). We modify the redshift dependence in Eq.~(\ref{eq:window_alignment_original}) by setting $\eta_{\rm IA}=0$ and using one amplitude parameter $A_{\mathrm{IA}, i}$ for each redshift bin $i$. This gives more flexibility to the modelling, at the cost of increasing the number of free parameters. Thus, the window function for the intrinsic alignment field reads:
\begin{align}\label{eq:window_alignment}
    W_i^{\rm I}(\chi) = - A_{\mathrm{IA}, i} \, \frac{C_1 \, \rho_{\rm cr} \, \Omega_{\rm m}}{D(\chi)} \, n_{i, \textrm{source}}(\chi) \ .
\end{align}

\begin{table}
  \centering 
  \renewcommand{\arraystretch}{2}
  \begin{tabular}{c c c c}
    &\textbf{S1}               &  \textbf{S2}  & \textbf{S3} \\
    \hline
    \hline
    $f_{\textrm{sky}}$ &  0.35 &  0.05 & 0.45 \\
    \hline
    $\bar{n}_{\textrm{source}}$ [gal/arcmin$^2$] & 30          &     50    & 50\\
    \hline
    $\bar{n}_{\textrm{lens}}$ [gal/arcmin$^2$] & 40          &     55    & 60\\
    \hline
    $\sigma_{\epsilon}$& 0.30                              &     0.25     & 0.35\\
    \hline
    \hline
    \end{tabular}
   \caption{Specifications for the three Stage IV surveys S1, S2 and S3 that we consider in this work. All details can be found in Sect.~\ref{sec:likelihoods}.}
  \label{tab:specs}
\end{table}

We include a multiplicative bias in our modelling of the cosmic shear signal. This keeps into account that the ellipticities measured in a survey are typically biased with respect to an ideal measurement, due to e.g. point spread function modelling errors, selection biases or detector effects \citep{Huterer06, Amara08, Kitching15,   Mandelbaum18, Taylor18, Mahony22, Cragg22}. We consider a multiplicative bias coefficient $m_i$ for each redshift bin $i$; therefore, the multiplicative bias rescales the cosmic shear power spectrum $C_{ij}^{\epsilon \epsilon}(\ell)$ by a factor $(1+m_i)(1+m_j)$. Another important source of uncertainty is the imperfect knowledge of the redshift distributions $n_{i, \textrm{source}}(z)$. Errors in these distributions have traditionally been modelled through the addition of a shift parameter $D_{z_i, \textrm{source}}$ for each redshift bin $i$, which changes the mean of the bin redshift distribution by shifting the distribution $n_{i, \textrm{source}}(z)$ to $n_{i, \textrm{source}}'(z) = n_{i, \textrm{source}}(z-D_{z_i, \textrm{source}})$ (see e.g. \citealp{Eifler21}).

\begin{figure*} 
    \includegraphics[width=0.84
\paperwidth]{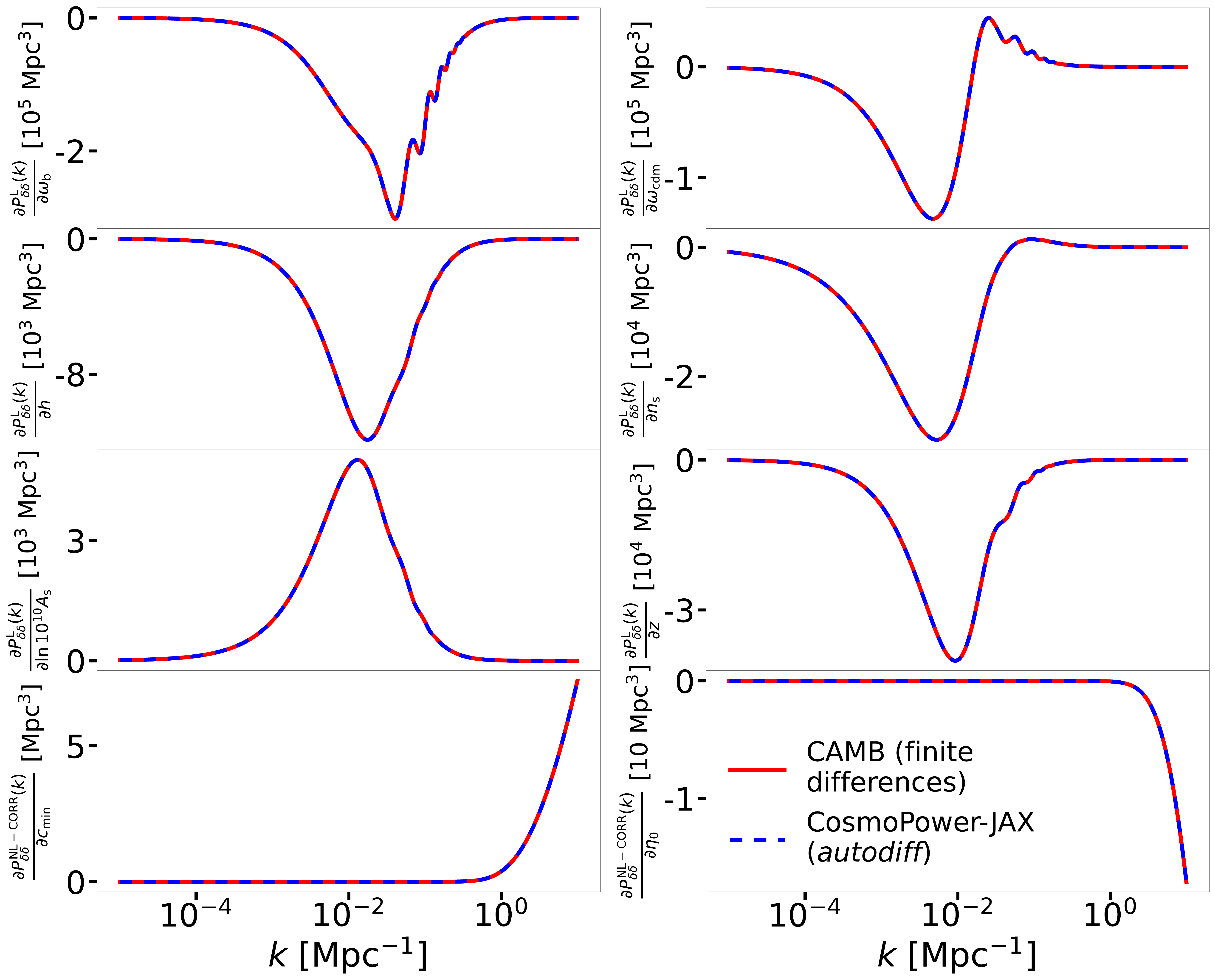}
\caption{Derivatives of the matter power spectrum components as defined in Eq.~(\ref{eq:pdd}) with respect to cosmological parameters for random samples in the test set. 
  Solid red lines represent finite differences calculated using CAMB and a five-point stencil as in Eq.~(\ref{eq:stencil}), while dashed blue lines are obtained with the \texttt{CosmoPower-JAX} emulator and automatic differentiation (\textit{autodiff}). A single derivative computed with finite differences requires $\mathcal{O} (10)$ s, as opposed to the $\mathcal{O} (0.1)$ s required for the derivatives with respect to all cosmological parameters obtained using \texttt{CosmoPower-JAX}.}
  \label{fig:acc_derivatives}
\end{figure*}

For the modelling of the galaxy clustering power spectrum $C_{ij}^{\mathrm{n} \mathrm{n}}(\ell)$ we assume a linear bias model between galaxy and dark matter density, with a free parameter $b_i$ for each redshift bin $i$. The window function $W_i^{\rm n}$ for the galaxy clustering field $\mathrm{n}$ is then given by:
\begin{align}\label{eq:window_clustering}
    W_i^{\rm n}(\chi) = b_i \, n_{i,\textrm{lens}}(\chi) \ ,
\end{align}
where for the galaxy clustering we use a different sample of redshift distributions $n_{i,\textrm{lens}}(\chi)$, which is also not perfectly known and therefore includes a set of shifts $D_{z_i, \textrm{lens}}$. We do not take into account redshift space distortions. 
Finally, the galaxy-galaxy lensing power spectrum is given by 
\begin{align}\label{eq:cell_galaxygalaxylensing}
C_{ij}^{\mathrm{n} \epsilon}(\ell) = C_{ij}^{\mathrm{n} \gamma}(\ell) + C_{ij}^{\mathrm{n} \mathrm{I}}(\ell) \ .
\end{align}

The first experiment we run is a simulated cosmic shear analysis of a Stage IV survey (dubbed ``S1''). We assume 10 tomographic bins for $n_{\textrm{source}}(z)$
, with galaxies following the distribution:
\begin{align}
    n(z) \propto z^2 \exp \left(- \left( z / z_0 \right)^{1.5} \right).
    \label{eq:nz}
\end{align}
with $z_0=0.64$ \citep{Smail94, Joachimi10}. We consider a Gaussian covariance matrix (e.g. \citealp{Tutusaus20}) with a sky fraction $f_{\textrm{sky}} = 0.35$, a surface density of galaxies $\bar{n}_{\textrm{source}}=30$ galaxies per arcmin$^2$, and an observed ellipticity dispersion $\sigma_{\epsilon}=0.3$. The $C(\ell)$ spectra are computed for 30 bin values between $\ell_{\textrm{min}}=30=$ and $\ell_{\textrm{max}}=3000$.

\begin{figure*}
    \includegraphics[width=0.836\paperwidth]{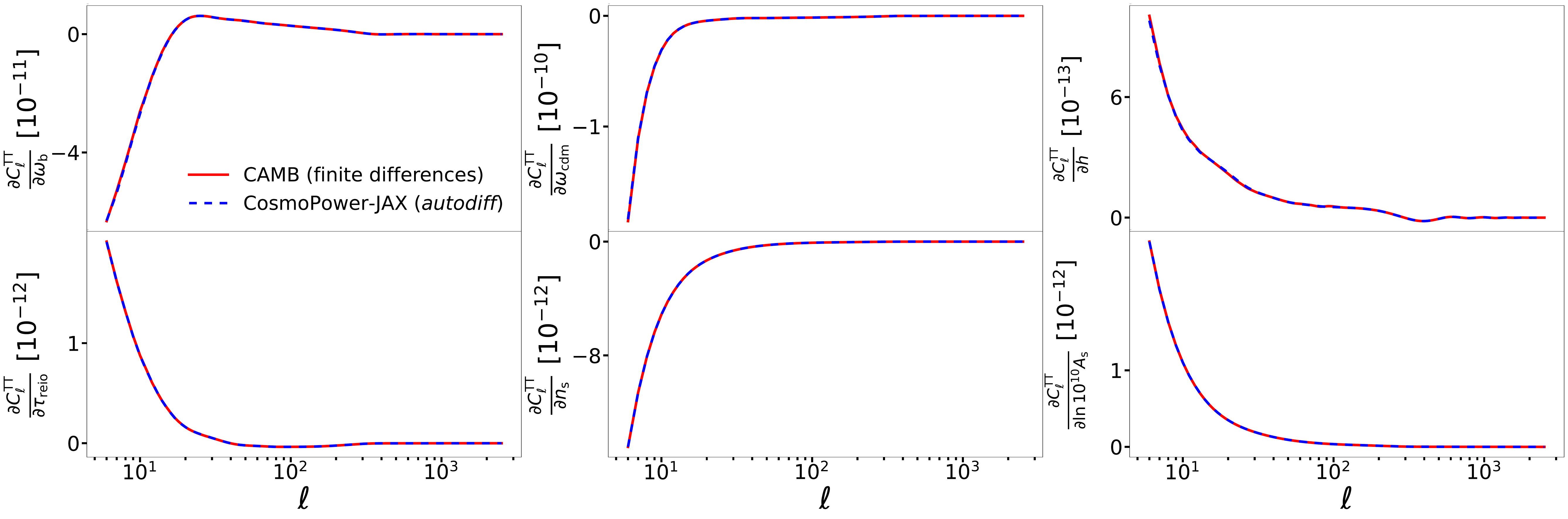}
  \caption{Same as in Fig.~\ref{fig:acc_derivatives} for the CMB temperature power spectrum $C_{\ell}^{\textrm{TT}}$.}
  \label{fig:acc_derivatives_tt}
\end{figure*}

\begin{figure*}
    \centering
    \includegraphics[trim={0.2cm 0 0.25cm 0},clip, scale=0.29]{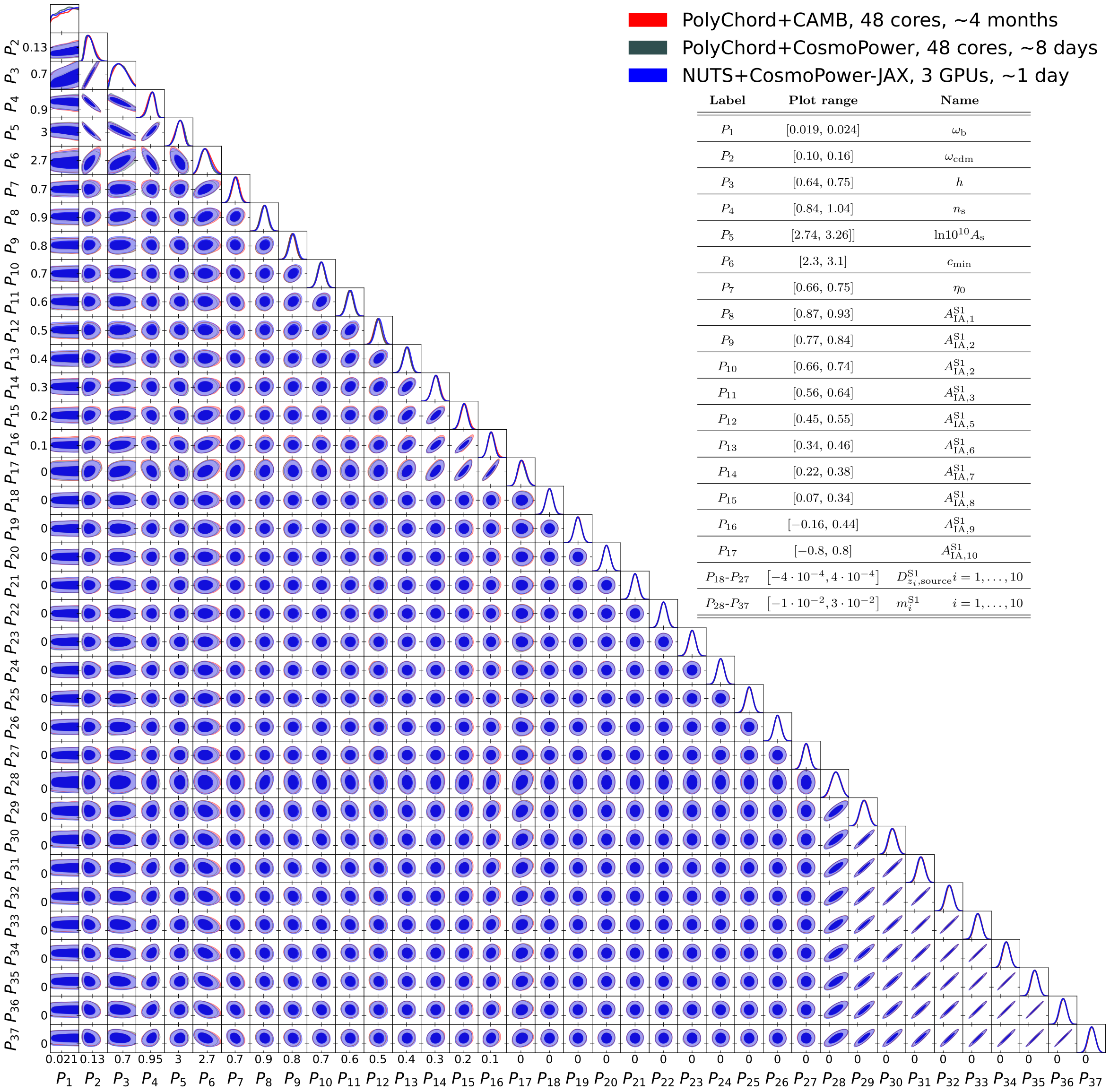}
  \caption{Inference results for the cosmic shear analysis of a simulated Stage IV survey with a total of 37 parameters. The contours show the 68\% and 95\% credibility contours and 1-D marginals for all of the parameters. The inset table reports the plot range and corresponding parameter name of every axis symbol; all parameters are described in Sect.~\ref{sec:likelihoods}, while in Table~\ref{tab:priors} we report the prior distributions and fiducial values. The contours in blue are obtained with the No U-Turn Sampler (NUTS) combined with our \texttt{CosmoPower-JAX} emulators. The contours in red are produced using the \texttt{PolyChord} sampler together with the Boltzmann code CAMB. The \texttt{CosmoPower-JAX} results are obtained over 3 graphics processing units (GPUs) in about 1 day, while the \texttt{PolyChord}+CAMB chain ran on 48 CPU cores for about 4 months, so that the total speed-up factor in terms of CPU/GPU hours is $\mathcal{O}(10^3)$. When replacing CAMB with the original \texttt{CosmoPower} emulator \citep{CP}, we obtain the contours in grey, which required about 8 days on 48 CPU cores, and show an excellent agreement overall. All of the contours shown in this paper are plotted using \texttt{GetDist} \citep{Lewis19}.}
  \label{fig:37_params}
\end{figure*}

We use the nested sampler \texttt{PolyChord} to sample the posterior distribution and run the inference pipeline within \texttt{Cobaya} \citep{Lewis13}. To compute the theoretical predictions for the cosmological observables we use the Core Cosmology Library (CCL, \citealp{Chisari19}). The three upper blocks of Table~\ref{tab:priors} summarise the prior distributions assumed for the parameters used in this analysis. These include:
\begin{itemize}
    \item five standard $\Lambda$CDM cosmological parameters, namely the baryon density $\omega_{\mathrm{b}} = \Omega_{\mathrm{b}} h^2$, the cold dark matter density $\omega_{\mathrm{cdm}} = \Omega_{\mathrm{cdm}} h^2$, the Hubble parameter $h$, the scalar spectral index $n_{\textrm{s}}$, and the primordial power spectrum amplitude $\mathrm{ln}10^{10}A_{\textrm{s}}$;
    \item the \texttt{HMcode} parameters $c_{\textrm{min}}$, $\eta_0$;
    \item one intrinsic alignment amplitude $A_{\textrm{IA}}$ for each of the ten redshift bins;
    \item one shift parameter $D_{z, \textrm{source}}$ for each of the ten source redshift distributions;
    \item one multiplicative bias coefficient $m$ for each of the ten redshift bins,
\end{itemize}
for a total of 37 model parameters.

We run a second experiment in which we consider the joint analysis of three surveys, dubbed ``S1'', ``S2'' and ``S3'', each one performing a 3x2pt analysis over 10 redshift bins over different $z$ ranges. $n_{\textrm{source}}(z)$ and $n_{\textrm{lens}}(z)$ are sampled from Eq.~(\ref{eq:nz}), with different galaxy densities $\bar{n}_{\textrm{source}}$ and $\bar{n}_{\textrm{lens}}$ reported in Table~\ref{tab:specs}. In this analysis, the cosmological and baryon parameters are unique, while each survey has its own set of nuisance parameters to model redshift distribution shifts, multiplicative biases and galaxy bias. With respect to the cosmic shear analysis, for each survey we add: 
\begin{itemize}
    \item one shift parameter $D_{z, \textrm{lens}}$ for each of the ten lens redshift distributions;
    \item one linear galaxy bias coefficient $b$ for each of the ten redshift bins,
\end{itemize}
for a total of 157 parameters. The prior distributions and fiducial values of the nuisance parameters for all surveys are reported in Table~\ref{tab:priors}, while we report all survey specifications in Table~\ref{tab:specs}.

\section{Results}\label{sec:results}
\subsection{Emulator accuracy}
Having used the same dataset and training settings, the accuracy of our \texttt{JAX} emulators is in excellent agreement with the results presented in SM22, i.e. the difference with respect to the CAMB test spectra is always sub-percent for both the matter power spectrum and the CMB probes. We do not show the accuracy plots here for brevity, and refer the reader to SM22 for an extended comparison.

In Fig.~\ref{fig:acc_derivatives} we then compare the derivatives of the linear matter power spectrum and the non-linear boost with respect to cosmological parameters, using CAMB and finite differences with a five-point stencil as a benchmark. These numerical derivatives rely on a simple Taylor expansion of the function $f(x)$ being considered, and are calculated as: 
\begin{equation}
    f'(x) \simeq \frac{-f(x+2s) + 8f(x+s) - 8f(x-s) + f(x-2s)}{12s} \ ,
    \label{eq:stencil}
\end{equation}
where $s$ is a small step size. We choose the step size for each derivative by ensuring that the numerical derivative would only change at the sub-percent level, or within machine precision, when increasing or decreasing $s$ by one order of magnitude. 

The derivatives computed by our \texttt{JAX} emulators are efficient to obtain and show excellent agreement with the numerical ones, despite the latter being prone to numerical instabilities due to the choice of the step size. A single derivative of the linear matter power spectrum with finite differences takes $\mathcal{O}(10)$ s on a CPU core, since it requires four independent calls to CAMB. On the other hand, on the same CPU core we can obtain derivatives with respect to all cosmological parameters and for batches of tens of spectra in $\mathcal{O}(0.1)$ s; we expect the speed-up to be even more substantial when running on GPU. In Fig.~\ref{fig:acc_derivatives_tt} we repeat this comparison for the $C_{\ell}^{\textrm{TT}}$ power spectrum, showing again good agreement and high computational efficiency for all cosmological parameters.

\subsection{Bayesian inference}
\label{sec:bayes_res}
Here we present and comment on the results of running our gradient-based inference pipelines. We emphasise that, while the survey configurations that we consider are representative of the ones expected from upcoming surveys, our goal is not to obtain official forecast predictions, but rather to focus on the challenge represented by the large parameter space required by these analyses, and how to implement viable solutions to efficiently explore these spaces.

In Fig.~\ref{fig:37_params} we show the posterior contours for the cosmological and nuisance parameters in the case of a cosmic shear analysis of a single Stage IV-like survey. The results obtained with NUTS and the \texttt{CosmoPower-JAX} emulators on three GPUs in one day are in excellent agreement with those yielded by the \texttt{PolyChord} sampler using CAMB and CCL, which however require about 4 months on 48 2.60 GHz Intel Xeon Platinum 8358 CPU cores. The total speed-up factor is then $\mathcal{O}(10^3)$ in terms of CPU/GPU hours, and $\mathcal{O}(10^2)$ in terms of actual elapsed time. We also run \texttt{PolyChord} with the original \texttt{CosmoPower} emulator (written using \texttt{TensorFlow}) on the same 48 CPU cores, obtaining overlapping contours in about 8 days. This suggests that a $\mathcal{O}(10)$ acceleration is provided by the neural emulator, with the NUTS sampler and the GPU hardware providing the remaining $\mathcal{O}(10)$ speed-up (in terms of the actual elapsed time).

\begin{figure}[b!]
    \centering
    \includegraphics[trim={0.3cm 0 0.2cm 0},clip, scale=0.45]{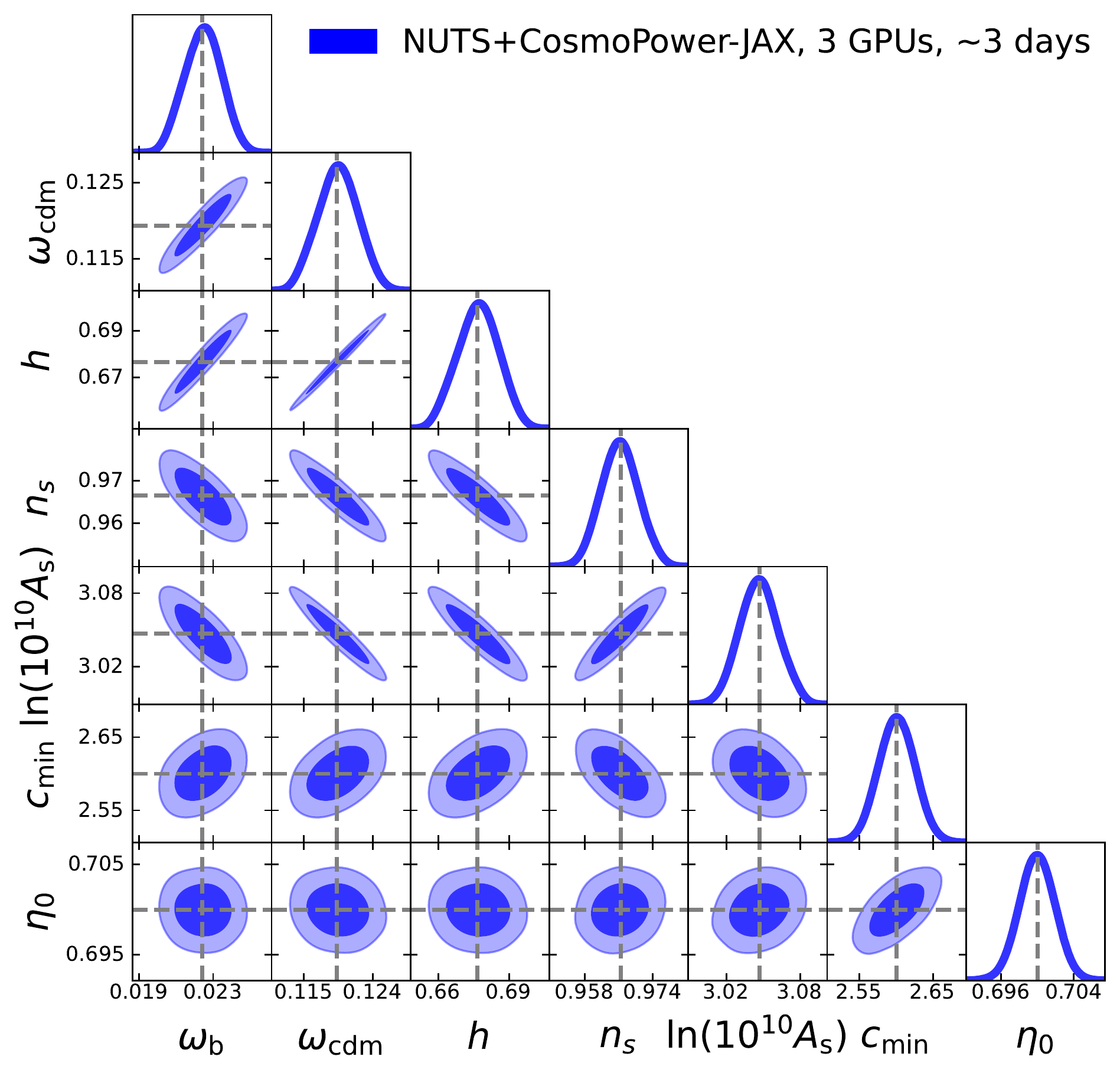}
  \caption{Inference results for the joint analysis of three simulated Stage IV surveys, each performing a 3x2pt analysis, for a total of 157 model parameters. Here we show the posterior contours for the cosmological parameters, while we report the marginal distribution of the nuisance parameters in Fig.~\ref{fig:157_params_nuisance}. The blue contours represent the 68\% and 95\% credibility contours and 1-D marginals. These are obtained with the No U-Turn Sampler (NUTS) combined with our \texttt{CosmoPower-JAX} emulators in about 3 days on 3 graphics processing units (GPUs). Dashed grey lines represent the fiducial values, reported in Table~\ref{tab:priors} together with the prior distributions.}
    \label{fig:157_params_cosmo}
\end{figure}%

\begin{figure*}
    \centering
    \includegraphics[scale=0.237]{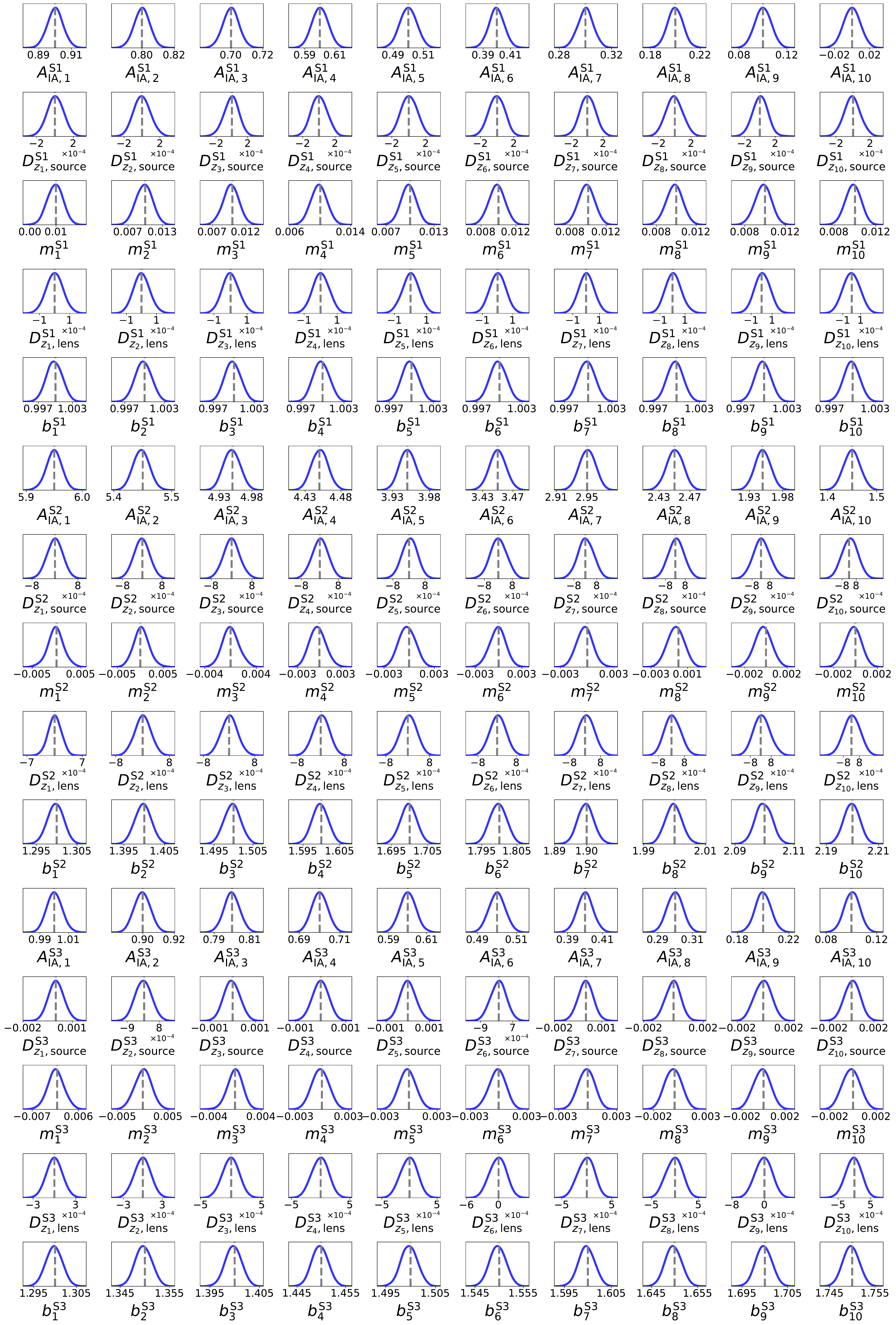}
  \caption{Marginal posterior distributions obtained in about 3 days on 3 graphics processing units (GPUs) with our \texttt{CosmoPower-JAX} emulators and the No U-Turn Sampler (NUTS) for the nuisance parameters of the the joint analysis of three simulated Stage IV surveys. Each survey performs a 3x2pt analysis, for a total of 157 parameters. The corresponding cosmological constraints can be found in Fig.~\ref{fig:157_params_cosmo}. Dashed grey lines represent the fiducial values, reported in Table~\ref{tab:priors} together with the prior distributions.}
  \label{fig:157_params_nuisance}
\end{figure*}

\begin{table}
  \centering 
  \renewcommand{\arraystretch}{2}
  \begin{tabular}{c c c c c c}
    &$\omega_{\mathrm{b}}$              &  $\omega_{\mathrm{cdm}}$  & $h$ & $n_{\textrm{s}}$ & $\mathrm{ln}10^{10}A_{\textrm{s}}$  \\
    \hline
    \hline
    \hline
    (a) & 0.02          &    $2\cdot 10^{-5}$   & $1\cdot 10^{-5}$ & $2\cdot 10^{-5}$ &  $4\cdot 10^{-5}$ \\
    (b) & 0.1          &    $2\cdot 10^{-5}$   & $1\cdot 10^{-5}$ & $2\cdot 10^{-5}$ &  $3\cdot 10^{-5}$ \\
    (c) &  0.02 &  0.02 & 0.02 & 0.03 & 0.03  \\
    \hline \hline
    (d) & 0.03          &    0.03   & 0.03 & 0.03 & 0.03 \\
    \hline
    \hline
    \end{tabular}
   \caption{Effective sample size per likelihood call (in percentage) $\eta$, as defined in Eq.~(\ref{eq:ess}), for: \newline(a) \texttt{PolyChord}+CAMB, cosmic-shear only; \newline(b) \texttt{PolyChord}+\texttt{CosmoPower}, cosmic-shear only; \newline(c) NUTS+\texttt{CosmoPower-JAX}, cosmic-shear only; \newline(d) NUTS+\texttt{CosmoPower-JAX}, joint analysis.\newline Except for $\omega_{\mathrm{b}}$, which is unconstrained in the case of the cosmic shear analysis, the combination of NUTS and \texttt{CosmoPower-JAX} yields values of $\eta$ higher by up to three orders of magnitude, confirming the higher efficiency of the sampling approach.}
  \label{tab:ess}
\end{table}

In Fig.~\ref{fig:157_params_cosmo} and Fig.~\ref{fig:157_params_nuisance} we then show the posterior contours for the 7 cosmological parameters and the 150 nuisance parameters, respectively, of the joint analysis of the three Stage IV-like surveys described in Sect.~\ref{sec:likelihoods}. Obtaining the corresponding contours using \texttt{PolyChord} and CAMB would be extremely computationally expensive: optimistically, assuming a linear (rather than exponential, see e.g. \citealt{Bishop95}) scaling of the run time with the number of inference parameters, we would expect to be able to obtain such contours in at least 6 years on 48 CPU cores. Therefore, we only show the results with NUTS and \texttt{CosmoPower-JAX}, which we obtain in about 3 days over 3 GPUs, for a total speed-up factor of $\mathcal{O}(10^4)$ in terms of CPU/GPU hours. These results are unbiased and in good agreement with the fiducial values of our analysis.

We further quantify the efficiency of the different sampling approaches by computing the effective number of samples per likelihood call, namely:
\begin{equation}
    \eta = \frac{n_{\textrm{eff}}}{n_{\textrm{eval}}} \ ,
    \label{eq:ess}
\end{equation}
where $n_{\textrm{eff}}$ is the effective sample size (ESS), computed with the \texttt{ArviZ} library \citep{Kumar19}, and $n_{\textrm{eval}}$ is the total number of likelihood calls. As we report and discuss in Table~\ref{tab:ess}, $\eta$ is almost always $\mathcal{O}(10^3)$ times bigger when using the NUTS sampler with \texttt{CosmoPower-JAX} with respect to the \texttt{PolyChord} sampler, and does not degrade even when considering 157 inference parameters.

\section{Conclusions}\label{sec:conclusions}

We presented \texttt{CosmoPower-JAX}, a \texttt{JAX}-based implementation of the cosmological neural emulator \texttt{CosmoPower} \citep{CP}. \texttt{CosmoPower-JAX} adopts the same emulation methods of the original \texttt{TensorFlow}-based version of \texttt{CosmoPower}, but implements them using the \texttt{JAX} library. The key feature of \texttt{JAX} is the ability to automatically differentiate functions written in common \texttt{Python} libraries, such as \texttt{NumPy}. Additionally, \texttt{JAX} allows \texttt{NumPy} software to be dynamically compiled, evaluated in batch, and run on graphics processing units (GPUs) and tensor processing units (TPUs). In this paper, we made use of these \texttt{JAX} features to show how they can accelerate Bayesian cosmological inference by orders of magnitude.

We started by training a set of \texttt{CosmoPower-JAX} emulators of matter and cosmic microwave background (CMB) power spectra, using the same datasets of \citet{CP}. We compared the accuracy of our \texttt{JAX}-based emulators with the Boltzmann code CAMB, showing excellent agreement both at the power spectra level as well as at the level of their derivatives. Crucially, rather than being obtained with finite differences, which are prone to numerical instabilities and require fine-tuning of the step size, derivatives of the power spectra with \texttt{CosmoPower-JAX} are obtained straightforwardly and efficiently with the built-in automatic differentiation (\textit{autodiff}) features provided by \texttt{JAX}. We note that the usefulness of our differentiable power spectra emulators is not limited to 2-point-statistics analyses of cosmological fields: for example, field-level inference pipelines \citep[e.g.][]{Makinen_2021, Makinen_2022, Loureiro23, Porqueres23} greatly benefit from efficient access to power spectra and their gradients.

We combined our emulators with the \texttt{JAX}-based cosmological library \texttt{jax-cosmo} \citep{Campagne23} to write likelihood functions that are fully differentiable by virtue of their pure-\texttt{JAX} implementation. Running these likelihoods within inference pipelines using gradient-based Monte Carlo algorithms like the No U-Turn Sampler (NUTS), we showed how to use \texttt{CosmoPower-JAX} to perform cosmological Bayesian inference over parameter spaces with $\mathcal{O}(100)$ parameters, using multiple graphics processing units (GPUs) in parallel. We performed a cosmic shear analysis for a Stage IV-like survey configuration with a total of 37 parameters, showing good agreement with standard nested sampling inference performed with CAMB, while being $\mathcal{O}(10^3)$ times faster. We finally showed that with \texttt{CosmoPower-JAX} and NUTS it is possible to perform a combined Bayesian inference on three different Stage IV-like surveys, each performing a joint cosmic shear and galaxy clustering analysis (3x2pt), for a total of 157 parameters, of which 150 represent nuisance parameters used to model systematic effects for each survey. We obtained unbiased posterior contours in about 3 days on 3 GPUs, as opposed to an optimistic estimate of at least 6 years needed to obtain the same results with traditional sampling methods and standard Boltzmann codes, for a total speed-up factor of $\mathcal{O}(10^4)$ in terms of CPU/GPU hours.

Developing fully differentiable, GPU-accelerated implementations of key analysis pipelines is of the utmost importance to tackle the massive computational requirements imposed by upcoming Stage IV surveys. To this purpose, we envision the use of differentiable libraries such as \texttt{TensorFlow}, \texttt{JAX} and \texttt{PyTorch} to become standard practice in cosmological software implementations, not least because they can all be used within the \texttt{Python} programming language, which is becoming the de facto standard programming language in both the cosmological and machine learning communities. 

Libraries such as \texttt{CosmoPower-JAX} and \texttt{jax-cosmo} are therefore of paramount importance for the success of next generation surveys. To this purpose, we plan to integrate the public release of \texttt{CosmoPower-JAX} with the \texttt{jax-cosmo} library. In parallel, we will explore improved gradient-based posterior samplers, such as those described in \citet{Steeg21, Park21, Robnik22, Wong2023}, to further improve the sampling efficiency, and thus the usefulness of our differentiable emulators.


\section*{Acknowledgements}

We are grateful to Erminia Calabrese, Benjamin Joachimi and Jason McEwen for useful discussions and feedback on this work. DP was supported by a Swiss National Science Foundation (SNSF) Professorship grant (No. 202671). ASM acknowledges support from the MSSL STFC Consolidated Grant ST/W001136/1. Part of the computations were performed on the Baobab cluster at the University of Geneva. This work has been partially enabled by funding from the UCL Cosmoparticle Initiative. The authors are pleased to acknowledge that part of the work reported on in this paper was performed using the Princeton Research Computing resources at Princeton University which is a consortium of groups led by the Princeton Institute for Computational Science and Engineering (PICSciE) and Office of Information Technology’s Research Computing.

\section*{Data Availability}

We make the \texttt{CosmoPower-JAX} emulators available in this GitHub repository (https://github.com/dpiras/cosmopower-jax, also accessible by clicking the icon \href{https://github.com/dpiras/cosmopower-jax}{\faicon{github}}).



\newpage
\bibliographystyle{mnras}
\bibliography{paper} 

\setlength{\twocolwidth}{\columnwidth}





\end{document}